\documentclass[prc,twocolumn,showpacs,showkeys,nofootinbib,superscriptaddress]{revtex4}
\usepackage{graphicx}% Include figurefiles
\usepackage{dcolumn}
\usepackage{epsfig}
\usepackage{bm}

\usepackage{amsfonts}
\usepackage{amsmath}
\usepackage{amssymb}
\usepackage[english]{babel}
\usepackage{graphicx}
\usepackage{epsfig}
\usepackage{bm}
\usepackage{verbatim} 
\usepackage[utf8]{inputenc}
\usepackage{booktabs}
\usepackage{multirow}
\usepackage{xcolor}
\usepackage[colorlinks=true,urlcolor=red,citecolor=red]{hyperref}
\usepackage[font=small]{caption}
\usepackage{float}
\usepackage{blindtext}

\newcommand{\sla}[1]{\displaystyle{\not} #1}

\begin{document}

\topmargin -1.20in

%\title{Quantum Field Theory of Neutrino oscillations in vacuum}
\title{Neutrino oscillations in Quantum Field Theory}
%\title{Neutrino oscillations in vacuum within quantum field theory}
%\title{Neutrino oscillations in vacuum in the framework of quantum field theory revisited}

\author{Sergey~Kovalenko}\email{sergey.kovalenko@unab.cl}
%\affiliation{Centro Cient\'{\i}fico-Tecnol\'{o}gico de Valpara\'{\i}so, Casilla 110-V, Valpara\'{\i}so, Chile}
\affiliation{Departamento de Ciencias F\'isicas, Universidad Andres Bello, 
  Sazi\'e 2212, Piso 7, Santiago, Chile}
\affiliation{ Millennium Institute for Subatomic Physics at the High Energy Frontier (SAPHIR), Fern\'andez
Concha 700, Santiago, Chile}
\author{Fedor~\v{S}imkovic}\email{fedor.simkovic@fmph.uniba.sk}
\affiliation{Faculty of Mathematics, Physics and Informatics, Comenius University in Bratislava, 842~48~Bratislava, Slovakia}
\affiliation{Institute of Experimental and Applied Physics, Czech Technical University in Prague, 128~00~Prague, Czech Republic}

\begin{abstract}
%{\color{blue}
We propose a Quantum Field Theory (QFT) approach to neutrino oscillations in vacuum. %based on quantum field theory  (QFT). 
The neutrino emission and detection are identified with the charged-current vertices of a single second-order Feynman diagram for the underlying process,   
%
%We identify the neutrino emission and detection to the charged-current vertices of a single second-order Feynman diagram for the process   
%$S+D\to \ell^{+}_{\alpha} +\ell^{-}_{\beta} + S' + D'$ 
enclosing neutrino propagation between these two points. 
The key point of our approach is the definition of the space-time setup 
typical for neutrino oscillation experiments, implying macroscopically large but finite volumes of the source and detector separated by a sufficiently large distance $L$. 
%For this setup 
We derive an $L$-dependent master formula for the charged lepton production rate, which provides the QFT basis for the analysis of neutrino oscillations.  Our formula depends on %hadronic part of 
the underlying  process and is not reducible to the conventional approach resorting to the concept of neutrino oscillation probability, which  originates from non-relativistic quantum mechanics (QM).  
We demonstrate that for some particular choice of  the underlying process our  QFT formula approximately coincides with the conventional one under some assumptions.
%originating from non-relativistic quantum mechanics (QM).  
%The corresponding neutrino oscillation probability contains a correction to the conventional QM expression, which are small only if neutrino masses are much less than their energy.  
%}
%This work has been supported by ANID PIA/APOYO AFB220004
%The field theory formalism of neutrino oscillations in a vacuum is presented. 
%The neutrino production and detection processes are part of a single Feynman diagram with initial and final states represented by plane waves. 
%The revised S-matrix approach guarantees the energy-momentum conservation in neutrino oscillations. The formalism is manifested in the process involving $\nu_\mu-\nu_e$ oscillations. A very good agreement with the conventional approach based on non-relativistic quantum mechanics is found.
\end{abstract}



\date{\today}

\maketitle

Three-quarters of a century ago Pontecorvo suggested the possibility of neutrino oscillations.
\cite{Pontecorvo:1957cp,Pontecorvo:1957qd}. Nowadays there exist clear evidences
for flavour neutrino oscillations from a variety of experimental
data on solar \cite{SNO:2002tuh}, atmospheric \cite{Super-Kamiokande:1998kpq},
reactor \cite{KamLAND:2002uet}, and accelerator neutrinos \cite{T2K:2013ppw}.
The existence of flavor conversion proves
%implies 
that neutrinos 
%mix with each other and 
have small, but non-zero masses, which in turn requires 
%particle 
a physics beyond the Standard Model. 
Neutrino oscillations are an interference phenomenon that allows the measurement of tiny quantities - neutrino masses.

%Neutrino oscillations are intrinsically a finite-time and finite-distance phenomenon, which can be decomposed in three steps:
%i) production of neutrinos in a source; ii) propagation of neutrinos to the detector;
%iii) interaction of neutrinos with a medium of the detector. 

\begin{figure}[t!]
 \includegraphics[width=0.9\linewidth]{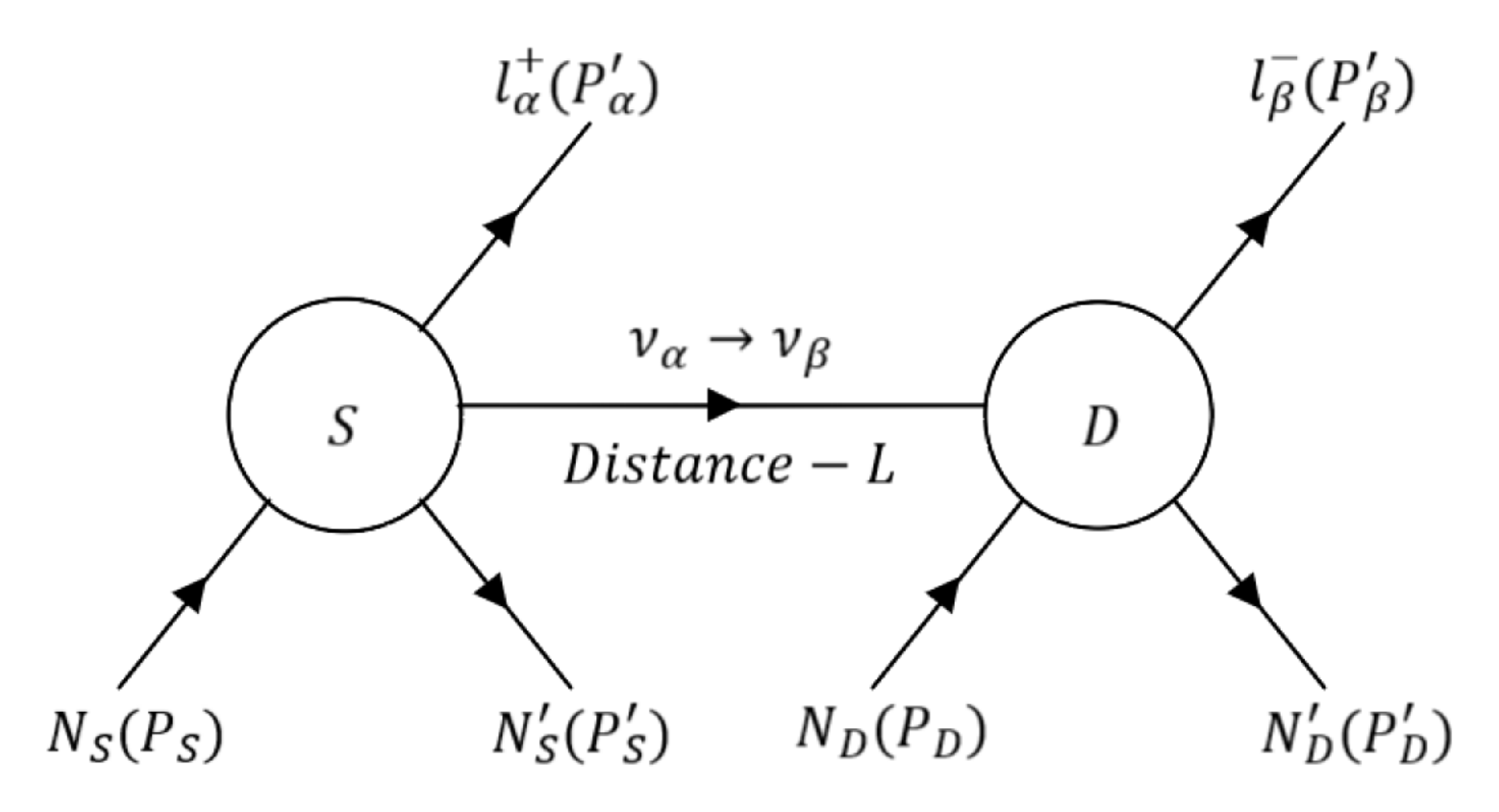}
 \caption{The Feynman diagram of the two subsequent charge-changing weak processes in source (S) and detector (D), separated by a macroscopical distance L. 
\label{fig:feynman}
}
 \end{figure}

%{\color{blue}
The typical neutrino oscillation experimental setup consists of a neutrino source {\bf S} and a detector {\bf D} separated by a distance $L$. The positive signal of neutrino oscillation 
$\nu_{\alpha}-\nu_{\beta}$ on the way from the source 
{\bf S} to the detector {\bf D} is the detection of the $\beta$-flavored charged lepton 
$\ell_{\beta}$  in response to the emission of the charged lepton $\ell_{\alpha}$ in the source {\bf S}. The latter indicates the emission of $\nu_{\alpha}$, created together with $\ell_{\alpha}$ in  the charged current (CC) decay of the source particle {\bf S}, as shown in Fig.~\ref{fig:feynman}.  
%
%The detector and the source have finite sizes and are separated by a finite distance, L. 
%
The neutrino oscillation data collected in these experiments are conventionally analyzed on the basis of  the quantum-mechanical concept of neutrino oscillation 
\cite{Gribov:1968kq,Bilenky:1976cw,Bilenky:1975tb,Bilenky:1978nj,Bilenky:1987ty,Bilenky:1980cx,Bilenky:1998dt,Bilenky:2008ez} introducing the oscillation probability
\begin{eqnarray}\label{eq:P-oscill-1}
\mathcal{P}_{\alpha\beta}(E_{\nu},L) &=& \left|\langle \nu_{\beta} | \nu_{\alpha}\rangle\right|^{2}=\\
\nonumber
&=&  \left|\sum^3_{j=1} U^*_{\alpha j} U_{\beta j} e^{-i m^2_j L/(2 E_{\nu})}\right|^2 
\end{eqnarray}
and considering the neutrino emission in {\bf S} and its detection in {\bf D} as three independent processes 
\begin{eqnarray}
\label{eq:Three-step-1}
S\to S' + \ell^{+}_{\alpha} + \nu_{\alpha},\ 
 \nu_{\alpha}\to \nu_{\beta},\   \nu_{\beta} + D \to  D' + \ell^{-}_{\beta}\, .
\end{eqnarray}
Therefore, we can write the total production rate of $\ell_{\beta}$ in the detectors as
\begin{eqnarray}
  \Gamma_{osc} = \int \frac{d \Phi_\nu(E_\nu)}{d E_\nu}
\frac{\mathcal{P}_{\alpha\beta}(E_{\nu},L)}{4\pi L^2}\sigma(E_\nu) dE_\nu \, .
\label{prodrnn}
\end{eqnarray}
Here, ${d \Phi_\nu(E_\nu)}/{d E_\nu}$ and $\sigma(E_\nu)$ are the energy distribution of (anti)neutrinos
from the source and cross-section for the scattering of neutrinos on the target at the detector, respectively. 
Here $(4\pi L^{2})^{-1}$ is the geometrical neutrino flux attenuation factor at the Detector site separated from the Source by the distance $L$. In the case of only one decaying particle, it describes the $L$-attenuation of the quantum mechanical probability corresponding to an outgoing  neutrino spherical wave.
In Eq.~(\ref{eq:P-oscill-1}) we have as usual
the  Pontecorvo-Maki-Nakagawa-Sakata (PMNS)
mixing matrix $U_{\alpha j}$ and  $E_{\nu}$ is the neutrino energy. 
%the energy $E_j = \sqrt{\vec{p}_{\nu}^{\, 2} + m_j^2},\, j=1,2,3$ of the neutrino with mass $m_j$ and momentum $\vec{p}_{\nu}$. 
%}
%
%The production rate is given by
%\begin{eqnarray}
%  R = \int \frac{d \Phi_\nu(E_\nu)}{d E_\nu}
%\frac{1}{4\pi L^2}\left\{\begin{array}{c}
%P_{\nu_\alpha \nu_\beta} \\
%P_{\overline\nu_\alpha \overline\nu_\beta}
%\end{array}\right\}
%\sigma(E_\nu) dE_\nu.\nonumber\\
%\label{prodrnn}
%\end{eqnarray}
%Here, ${d \Phi_\nu(E_\nu)}/{d E_\nu}$ and $\sigma(E_\nu)$ are the energy distribution of (anti)neutrinos
%from the source and cross-section for scattering of neutrinos on target at the detector, respectively. 
%The production of (anti)neutrinos with flavor $\alpha$ at neutrino source and following
%detection of (anti)neutrinos with flavor $\beta$ ($\alpha = e,~\mu$, and $\tau$) at detector
%are weighted with the probability of (anti)neutrino oscillations,
%\begin{eqnarray}
%  P_{\nu_\alpha \nu_\beta} &=& \left|\sum^3_{j=1} U^*_{\alpha j} U_{\beta j} e^{-i m^2_j L/(2 E_j)}\right|^2, \nonumber\\
%  P_{\overline\nu_\alpha \overline\nu_\beta} &=& \left|\sum^3_{j=1} U_{\alpha j} U^*_{\beta j} e^{-i m^2_j L/(2 E_j)}\right|^2.
%\label{pnunu}
%\end{eqnarray}
%Here, $U_{\alpha j}$ are elements of Pontecorvo-Maki-Nakagawa-Sakata (PMNS) neutrino
%mixing matrix and masses of neutrinos. $E_j$ is the energy of the neutrino with mass $m_j$ and momentum p
%($E_j = \sqrt{p^2 + m_j^2}$, j=1,2, and 3). L is the distance between the neutrino source and the detector. 
%

In the three neutrino oscillation global fits to neutrino oscillation data, six parameters are determined, in particular, the two differences of squared neutrino masses  (one of which can be positive or negative), 3 mixing angles, and the Dirac CP phase \cite{deSalas:2017kay,Capozzi:2017ipn,Esteban:2020cvm}. 
%{\color{blue} 
This analysis used the  QM approach based on Eqs. (\ref{eq:P-oscill-1})-(\ref{prodrnn}) and implying plane wave representation for the oscillating  neutrinos.  The theoretical difficulties of such an approach have long been known. In view of these difficulties, several improvements to the QM plane wave approach have been proposed in the literature.
%On the other hand, neutrino oscillations is a phenomenon, to which non-relativistic approximation is not always applicable.  
%
%The need for a firm theoretical basis of the interpretation of the oscillation data of these parameters 
%Reliable extraction of these fundamental parameters from neutrino oscillation data requires a solid theoretical foundation as a relativistic quantum phenomenon within the quantum field theory, which can be applied to data analysis without additional assumptions from the theory side. 
%
%The theory of neutrino oscillations remains to be an open problem.
%Considering the importance of the neutrino oscillation analysis, formulation of a robust theoretical framework describing this phenomenon is attracting significant efforts.
%
%continues to be a subject of great interest. 
%However, its consistent description as a relativistic quantum phenomenon in the framework of quantum field theory is still missing.
%
%It still lacks a self-consistent description as a relativistic quantum phenomenon within the quantum field theory, which can be applied to data analysis without additional assumptions from the theory side. 
%There is a means that neutrino oscillations contradict the standard treatment of oscillations. 
%
%model 
One of them resorts to the representation of oscillating neutrinos in the form of wave packets \cite{Kayser:1981ye,Kobzarev:1981ra,Giunti:1991sx,Giunti:1993se,Beuthe:2002ej,Beuthe:2001rc,Kayser:2010bj}. This
idea was further developed in a series of papers
\cite{Akhmedov:2010ms,Akhmedov:2010ua,Akhmedov:2012uu,Naumov:2013uia,Akhmedov:2017mcc,Naumov:2020yyv,Falkowski:2019kfn,Cheng:2022lys,Naumov:2022kwz}.
Although the quantum mechanical wave-packet approach is a significant improvement to the standard plane-wave treatment, it also suffers from drawbacks. Among them we may mention:
 ill-defined flavor states and difficulties with the determination of the wave packet size.
On the other hand, both approaches has limited applicability to neutrino oscillations. This fact raises the question about the possibility of description of neutrino oscillation as a relativistic quantum phenomenon in the framework of Quantum Field Theory (QFT).
%
%The theory of neutrino oscillations remains to be an open problem. 
A variant 
%In parallel with the wave packet approach, 
of the QFT approach with plane waves has been developed in Refs. \cite{Grimus:1998uh,Grimus:1999ra,Grimus:2019hlq}, where the neutrino production and detection processes are assigned to be  part of a single Feynman diagram, and the corresponding decay rate and cross section are computed in the standard way.
This approach is free from the ambiguities of the standard QM plane wave treatment related to the equality of the energies or momenta of the different mass eigenstates and the choice of the reference frame. 
%
%This approach avoides the ambiguities of the standard non-relativistic plane wave treatment, namely the question about the equality of the energies or momenta of the different mass eigenstates and the proper choice of the reference frame. 
%There is a belief that the QFT framework might provide a more realistic interface between theory
%and experiment once it is properly formulated. 
%
%{\color{blue}
%There is a belief that the QFT framework can provide a more realistic interface between theory
%and experiment once the problem is properly formulated and treated. 
%}

However, the following important question remains open:
Can neutrino oscillations be consistently described in the standard QFT S-matrix formalism, and if not, how should this formalism be adapted to make the description possible?
%
%Can neutrino oscillations be consistently described in the standard S-matrix formalism of QFT, and if not, how should this formalism be adopted to make the description possible? 
%
This and related questions have been addressed, for instance, in 
 %in a number of publications 
 Refs.~\cite{Egorov:2017qgk, Volobuev:2017izt,Egorov:2019vqv}. 
The goal of the present  paper is to show that neutrino oscillations can be consistently described within the QFT framework, if the space-time setup of neutrino oscillations is properly modeled for the application of the S-matrix approach. 

%, if it is properly done. 

%
%
%We assume two subsequent two weak processes in a source and at detector as follows (see Fig. \ref{fig:feynman}):
%\begin{eqnarray}\label{genprocess}
%N_S(P_S) &\rightarrow& N'_S(P'_S) + l^+_\alpha(P'_\alpha) + \nu_k(P_k), \nonumber\\
%\nu_k(P_k) + N_D(P_D) &\rightarrow& N'_D(P'_D) + l^-_\beta(P'_\beta). 
%\end{eqnarray}
%
In our QFT approach the setup in Fig.~\ref{fig:feynman} is a Feynman diagram describing the process 
\begin{eqnarray}\label{eq:FeynDiagr-1}
S+D\to \ell^{+}_{\alpha} +\ell^{-}_{\beta} + S' + D'.
\end{eqnarray}
Here, $S(D)$  and $S'(D')$ represent initial and final hadrons or nuclei in the vertex 
{\bf S}({\bf D}), respectively. Of course, if a meson decay is considered,
$S'$ hadron is missing. The flavor indices $\alpha, \beta$ stand for $e, \mu, \tau$.
%$l^\pm_\alpha$ with $\alpha = e, \mu \tau$ stands for $e^\pm$, $\mu^\pm$, and $\tau^\pm$, respectively. 
The 4-momenta of hadrons and leptons 
%and neutrino 
are as follows: $P_{S,D} \equiv (E_{S,D},\mathbf{p}_{S,D})$,
$P'_{S,D} \equiv (E'_{S,D},\mathbf{p'}_{S,D})$ and $P_{\alpha, \beta} \equiv (E_{\alpha, \beta},\mathbf{p}_{\alpha, \beta})$.
%$P_{k} \equiv (E_{k},\mathbf{p}_{k})$ with k=1, 2, and 3. 
%
%Here, $N$  and $N'$ represent initial and final hadrons (or nuclei), respectively. Of course, if meson decay is considered,
%$N'$ hadron is missing. $l^\pm_\alpha$ with $\alpha = e, \mu \tau$ stands for $e^\pm$, $\mu^\pm$, and $\tau^\pm$, respectively. 
%The 4-momenta of hadrons, leptons and neutrino are as follows: $P_{S,D} \equiv (E_{S,D},\mathbf{p}_{S,D})$,
%$P'_{S,D} \equiv (E'_{S,D},\mathbf{p'}_{S,D})$, $P_{\alpha, \beta} \equiv (E_{\alpha, \beta},\mathbf{p}_{\alpha, \beta})$,
%$P_{k} \equiv (E_{k},\mathbf{p}_{k})$ with k=1, 2, and 3. 
%
%For simplicity, energies of neutrinos are considered to be significantly below the mass of the $W^\pm$ boson.
%Both, the production and annihilation of neutrinos are described with the V-A effective Hamiltonian:
%\begin{eqnarray}
%  H_{V-A} = \frac{G_F}{\sqrt{2}} J^\mu J_\mu^\dagger~~~{\rm with}~~~
%J^\mu &=& j^\mu_{\rm had} + j^\mu_{\rm lep}, \nonumber\\
%\end{eqnarray}
%where the charged weak hadronic and leptonic currents are given by
%\begin{eqnarray}
%j^\mu_{\rm had} &=&
%  \left(\overline{u}, \overline{c}, \overline{t}\right) \gamma^\mu\left(1-\gamma_5\right) V_{\rm CKM}
%\left(\begin{array}{c}
%  d \\
%  s\\
%  b
%\end{array}\right)\nonumber\\
%j^\mu_{\rm lep} &=& \left(\overline{\nu}_1, \overline{\nu}_2, \overline{\nu}_3\right)
%\gamma^\mu\left(1-\gamma_5\right) U_{\rm PMNS}
%\left(\begin{array}{c}
%  e \\
%  \mu\\
%  \tau
%  \end{array}\right).
%\end{eqnarray}
%Here, $V_{\rm CKM}$ is the Cabbibo-Kobayashi-Maskava quark mixing matrix and $U_{\rm PMNS} = U$. 
%
In 
%the Lorentz covariant perturbation theory
the S-matrix approach 
%the matrix element of the
 the process (\ref{eq:FeynDiagr-1}) appears in the second order in the Charged Current (CC) interactions 
\begin{eqnarray}\label{eq:CC-1}
\mathcal{L}^{CC}&=& - 2\sqrt{2}G_{F}\, \overline{\ell_{\alpha}} \gamma^{\mu}P_{L} U_{\alpha i} \nu_{i} \cdot J_{\mu},   
\end{eqnarray}
where $J_{\mu}= \bar{u}_{i}\gamma_{\mu} P_{L} V^{CKM}_{ij}d_{j}$ is the hadronic current and $\nu_{i}$ is the neutrino state with mass $m_{i}$.
The CC vertices, located in 
{\bf S} and {\bf D}, are connected by the neutrino propagator. 
The S-matrix element
%, which is a transition $i\to f$ amplitude $S_{if}$, 
corresponding to the diagram  in Fig.~\ref{fig:feynman} is given by
%
%Within the S-matrix approach the matrix element of the compound process (\ref{genprocess})
%in the second order perturbation theory in $G_F$ takes the form.
\begin{eqnarray}\label{ampl}    
&&  \langle f|S^{(2)}|i\rangle
= - i \int d^4 x_1 J^\mu_{S}(P'_{S},P_{S}) e^{i(P_\alpha + P'_{S} - P_{S})\cdot x_1} \times \nonumber\\
&&  \int d^4 x_2 J^\mu_{D}(P'_{D},P^{}_{D}) e^{i(P_\beta + P'_{D} - P^{}_{D})\cdot x_2}
\sum_{k=1}^3 U^*_{\alpha k} U^{}_{\beta k}  \times \\
\nonumber
&& \overline{v}(P_\alpha;\lambda_{\alpha})\gamma_\mu(1-\gamma_5) D(x_2-x_1,m_k)
(1-\gamma_5)\gamma_\nu u(P_\beta;\lambda_{\beta})
  %  \int \frac{d^4 p}{(2\pi)^4} e^{-i p\cdot (x_1-x_2)}
\end{eqnarray}
where $\lambda_{\alpha,\beta}$ are the polarizations of the final leptons and
\begin{eqnarray}
  J^\mu_{I}(P'_{I},P^{}_{I}) e^{i(P'_{I} - P^{}_{I})\cdot x} \equiv
  \sqrt{2}\,  G_F \langle I' | J^{\mu} (x)| I \rangle
%
%  J^\mu_{I}(P'_{I},P^{}_{I}) e^{i(P'_{I} - P^{}_{I})\cdot x} \equiv
%  \frac{G_F}{\sqrt{2}} \langle N'_{I}|j^\mu_{\rm had}(x)|N_{I}\rangle
\end{eqnarray}
%are the expectation values of hadronic currents (I=S,D).
is the definition for normalized hadronic current matrix elements with $I=S,D$.

The amplitude (\ref{ampl}) contains the contributions of all the possible space-time configurations with arbitrary distance $\mathbf{x}_{2}-\mathbf{x}_{1}$ between the particles in the source {\bf S} and in the detector {\bf D} and arbitrary time ordering $t_{1}>t_{2}, t_{1}<t_{2}$  of  the moments $t_{1} =x^{0}_{1}$ and $t_{2}=x^{0}_{2}$ of the charged fermions 
$\ell^{+}_{\alpha}$ and $\ell^{-}_{\beta}$  emission, respectively. 

The key point of our approach is the definition of the space-time settings
 %of the charged lepton emission in the source and detector, 
 taking place in any long-baseline  neutrino oscillation experiments:\\
 {\bf (a)} The source {\bf S} and detector {\bf D} have 
 macroscopically large but finite volumes  $V_{S,D}\sim \ell_{S,D}^{3}$ with the characteristic sizes $\ell_{S,D}$;\\
 {\bf (b)} the  {\bf S} and {\bf D} are separated by a distance  $L$ such that
 \begin{eqnarray}\label{eq:finite-size}
&& \ell_{S,D} \ll L
\end{eqnarray}
{\bf (c)}  In view of (\ref{eq:finite-size}) one can experimentally distinguish the event when the charged leptons $\ell_{\alpha}^{+}$ and 
$\ell_{\beta}^{-}$ are emitted at $(\mathbf{x}_{1}, t_{1})$ and $(\mathbf{x}_{2}, t_{2})$, where 
$\mathbf{x_{1,2}}\in V_{S,D}$, respectively. Therefore, the space integrations in Eq.~(\ref{ampl}) over  $\mathbf{x}_{1}$ and  $\mathbf{x}_{2}$   are limited to the volumes $V_{S}$ and $V_{D}$ of the source 
{\bf S} and the detector {\bf D}, respectively.\\
{\bf (d)} Since the emission of $\ell_{\beta}^{-}$ is caused by the absorption of the neutrino emitted in the source, the time ordering $t_{1}<t_{2}$ is required. 
%Since the emission of $\ell_{\beta}^{-}$ is caused by the absorption of the neutrino emitted in the source, the time ordering $t_{1}<t_{2}$ is required. 
%} 

The contribution with the correct time ordering are readily extracted using
the fact that the T-ordering is controlled by the neutrino propagator,  
which can be represented as 
\begin{eqnarray}\label{eq:adv-ret-1}   
%\nonumber
%
  D(x;m) &=& \theta (x_{0}) D^{-}(x;m) + \theta(-x_{0}) D^{+}(x;m), 
\end{eqnarray}
where the retarded $D^{+}(x)$ and advanced  $D^{-}(x)$ propagators are
\begin{eqnarray}\label{eq:adv-ret-2}
&& D^{\pm}(x;m) = \nonumber\\
&&~~  \int \frac{d \mathbf{q} }{(2\pi)^{3}}  \frac{\mp(-\mathbf{q} \cdot \bm{\gamma} + \omega \gamma^{0}) + m}{2 \omega} e^{\pm i (-\mathbf{q}\cdot\mathbf{x}  + \omega x_{0}) }
\end{eqnarray}
with $\omega =\sqrt{\mathbf{q}^{\, 2} + m^{2}}$. 
%
%Neutrino oscillations are intrinsically
%a finite-time and finite-distance phenomenon. The time-space locations of the weak interactions
%at the source and detector are $x_1 = (t_S, \mathbf{x}_1)$  and $x_2 = (t_D, \mathbf{x}_2)$,
%respectively, i.e., $t_D > t_S$. 
%{\color{blue}
Replacing in (\ref{ampl}) the neutrino propagator 
$D(x_{2}-x_{1};m)\to  \theta (x^{0}_{2}-x^{0}_{1}) D^{-}(x_{2}-x_{1};m)$ we explicitly carry out the integration over $x^{0}_{1}$ and $x^{0}_{2}$. This results in
%}
%The integration over time-variables results in 
%
the product of the delta function reflecting energy conservation in the process (\ref{eq:FeynDiagr-1}) and the energy denominator:
\begin{eqnarray}
  2\pi i \frac{\delta(E_\beta + E'_D - E_D + E_\alpha + E'_S - E_S)}
  {\omega + E_\alpha + E'_S - E_S + i \varepsilon}.
\end{eqnarray}
%The integral over $\mathbf{q}$ is calculated by the theory of residuals
%{\color{blue}
The integration over $\mathbf{q}$ can also be carried out explicitly by integrating over the angles and analyzing the analytic structure of the integrand in the complex $|\mathbf{q}|$-plane. In this way we find
% is calculated by the theory of residuals
%}
\begin{eqnarray}
\label{eq:q-int-1}
&&  \int \frac{d \mathbf{q} }{(2\pi)^{3}}
  \frac{\sla{q} + m_k}
       {2 \omega (\omega + E_\alpha + E'_S - E_S + i\varepsilon)}
       e^{i \mathbf{q}\cdot(\mathbf{x}_2-\mathbf{x}_1)}\nonumber\\
&\simeq& \frac{1}{4\pi}
\frac{e^{i {p}_k {|\mathbf{x}_2 - \mathbf{x}_1|}}}{|\mathbf{x}_2-\mathbf{x}_1|} \left( \sla{Q}_k + m_k\right). 
\end{eqnarray}
Here, $Q_k \equiv (E_\nu, \mathbf{p}_k)$,
\begin{eqnarray}
  E_\nu = E_S -  E'_S - E_\alpha = E_\beta + E'_D - E_D,
  \label{nuencon}
\end{eqnarray}
$\mathbf{p}_k = p_k \left(\mathbf{x}_2-\mathbf{x}_1\right) /|\mathbf{x}_2-\mathbf{x}_1|$
and $p_k = \sqrt{E^2_\nu - m^2_k}$.  
%{\color{blue} 
We note that according (\ref{nuencon}), energy conserves separately in each vertex of the diagram in Fig.~\ref{fig:feynman}, which correspond to the CC processes in  the source and detector.
%
%We note the presence of separate energy conservations associated with both weak processes, in the source and detector, through the definition of neutrino
%energy $E_\nu$ in Eq. (\ref{nuencon}).
%}

%{\color{blue}
Now, we adjust the remaining $\mathbf{x}_{1,2}$-integration to the space-time setup 
{\bf (a)-(c)} specified at (\ref{eq:finite-size}). 
To this end we introduce the reference points inside the source $O_{S}$ and the detector $O_{D}$. The vectorial distance between $O_{S}$ and $O_{D}$ we denote by 
$\mathbf{L}$, with $|{\bf{L}}|=L$, introduced in (\ref{eq:finite-size}).  Since the integrations over  
$\mathbf{x}_{1,2}$ run inside the volumes $V_{S,D}$, we replace the variables 
$(\mathbf{x}_{1},  \mathbf{x}_{2}) \to (\mathbf{x}_{S},  \mathbf{x}_{D}+\mathbf{L})$.

%The detector and the source have finite sizes and are separated by a finite distance L.
%We introduce the position of the first interaction from the center of source $\mathbf{x}_S$
%and the position of the second interaction from the center of detector $\mathbf{x}_D$. We have
%$\mathbf{x}_2-\mathbf{x}_1  = \mathbf{x}_D + \mathbf{L} - \mathbf{x}_S$
%where $L=|\mathbf{L}|$ is the distance between the source and the detector. 
As a consequence of (\ref{eq:finite-size}) and, therefore, $L \gg |\mathbf{x}_S|$, $L \gg |\mathbf{x}_D|$, we get 
%}
%with good approximation
\begin{eqnarray}
\label{eq:approx-q}
  \frac{e^{i {p}_k {|\mathbf{x}_2 - \mathbf{x}_1|}}}{|\mathbf{x}_2-\mathbf{x}_1|} \simeq
  e^{i \mathbf{p}_k \cdot \mathbf{x}_D}~ e^{-i \mathbf{p}_k \cdot \mathbf{x}_S}~\frac{e^{i p_k L}}{L}.
\end{eqnarray}
%We note that conventionaly a single finite volume is assumed for the source-detector system.
%However, being both sufficiently large and separated by macroscopically large distance L,
%in the presented formalism of neutrino oscillations  two separate volumes  $V_S$ and $V_D$
%related, respectively, to the source and detector are considered. 
%{\color{blue} 
Inserting  (\ref{eq:adv-ret-1})-(\ref{eq:approx-q}) to Eq.~(\ref{ampl}) and integrating over $\vec{x}_{S,D}$ we find
%
%The particles states
%associated with the weak interaction located in the source (detector) are normalized to the volume
%$V_S$ ($V_D$). 
%By inserting above components into Eq. (\ref{ampl}), performing substitutions
%$\mathbf{x}_1=\mathbf{x}_S$,  $\mathbf{x}_2=\mathbf{x}_D + \mathbf{L}$ and
%integration over $\mathbf{x}_S$ and $\mathbf{x}_D$, we obtain  
%
\begin{eqnarray}\label{ampl2}
&&\langle f|S^{(2)}|i\rangle = 
%
%(2\pi)^4\delta(P_\beta + P'_D - P_D + P_\alpha + P'_S - P_S) \times \nonumber\\
%%
%&&~~~~~e^{-i(\mathbf{p}_\beta + \mathbf{p}'_D - \mathbf{p}^{}_D)\cdot \mathbf{L}}~ J^\mu_{S}(P'_{S},P^{}_{S}) J^\mu_{D}(P'_{D},P^{}_{D})\times\nonumber\\  
%&&~~~~~ \sum_k U_{\alpha k} U^*_{\beta k}~\frac{e^{i p_k L}}{4\pi L}~ 
%(2\pi)^3 \delta(\mathbf{p}_k + \mathbf{p}_\alpha + \mathbf{p}'_S - \mathbf{p}_S)\times \nonumber\\
%&&~~~~~~~~~~~ \overline{u}(P_\alpha)\gamma_\mu(1-\gamma_5) \left( \sla{P}_k + m_k\right)
%(1-\gamma_5)\gamma_\nu v(P_\beta)\nonumber\\
%2\pi \delta(E_\beta + E'_D - E_D + E_\alpha + E'_S - E_S)\times\\
(2\pi)^{7} \delta(E_{f}-E_{i}) \sum_k  U_{\alpha k} U^*_{\beta k}~\frac{e^{i p_k L}}{4\pi L}
%\langle f|T_k|i\rangle 
\times\\
\nonumber
&&
%(2\pi)^6 
T^{\alpha\beta}_{k}\delta^{3}_{V_{S}}(\mathbf{p}_k + \mathbf{p}_\alpha + \mathbf{p}'_S - \mathbf{p}_S)
~\delta^{3}_{V_{D}}(\mathbf{p}_\beta + \mathbf{p}'_D - \mathbf{p}_D - \mathbf{p}_k)\, ,
%\nonumber\\
\end{eqnarray}
where $E_{f}-E_{i} = E_\beta + E'_D - E_D + E_\alpha + E'_S - E_S$ and 
\begin{eqnarray}
%\nonumber
\label{eq:def-T-1}
&&
%\langle f| T_{k}|i\rangle
T^{\alpha\beta}_{k} =
%
%e^{-i(\mathbf{p}_\beta + \mathbf{p}'_D - \mathbf{p}^{}_D)\cdot \mathbf{L}}~ 
%
J^\mu_{S}(P'_{S},P^{}_{S}) J^{\nu}_{D}(P'_{D},P^{}_{D})  \times
%&& U_{\alpha k} U^*_{\beta k}~\frac{e^{i p_k L}}{4\pi L}~ 
% \\
% \nonumber
%&& 
\\
\nonumber
&&\overline{v}(P_\alpha;\lambda_{\alpha})\gamma_\mu(1-\gamma_5)\sla{Q}_k \gamma_\nu u(P_\beta;\lambda_{\beta}).
\end{eqnarray}
%{\color{blue} 
Here we have introduced the finite-volume delta function
$\delta^{3}_{V}(\mathbf{p})$, which tends to the usual one 
$\delta^{3}_{V}(\mathbf{p})\to \delta^{3}(\mathbf{p})$
%$\delta^{3}_{V}(\mathbf{p})\to \delta^{3}(\mathbf{p})$ 
in the limit $V\to \infty$. Thus, the 3-momentum in the vertices of the diagram in Fig.~(\ref{fig:feynman})  is conserved only approximately. However, since the volumes $V_{S,D}\sim \ell_{S,D}^{3}$ are macroscopically large, a deviation $\Delta P \sim \ell_{S,D}^{-1}$ from the 3-momentum conservation is negligibly small and has no practical significance.
%compared to the momenta 
%$\vec{P}_{k}$ appearing in both vertices of the diagram Fig.~\ref{fig:feynman}.
%in the process (\ref{eq:FeynDiagr-1}) 
%in the rest-frames of  the reference points $O_{S,D}$ defined below (\ref{eq:q-int-1}).
Therefore, in the final results we use 
$\delta^{3}_{V_{S,D}}(\mathbf{p})\approx \delta^{3}(\mathbf{p})$ as a good approximation.
%}

%{\color{blue}
%It is worth mentioning that unlike in the the standard QM approach there is no factorization of the process
%of production, propagation and absorption of neutrinos. These processes are coupled by the lepton current
%in Eq. (\ref{eq:def-T-1}) as QFT neutrino propagator is introduced. The factorization can not be achieved by
%a transformation
%\begin{eqnarray}
%&&(1-\gamma_5)~ \sla{Q}_k = \\
%&&\frac{1}{2}~ (1-\gamma_5)~ \sum_{\lambda_k} u(Q_k;\lambda_k) \overline{u}(Q_k;\lambda_k)~ 
%(1+\gamma_5)\nonumber
%\end{eqnarray}  
%as neutrinos are massive particles and there is a sum over both polarizations of neutrino.
%Consequently, the trace corresponding to the lepton current in Eq. (\ref{eq:def-T-1})
%has to be calculated.
%}

%
With this at hand we derive the  central result of the present paper, which is the {\it Master Formula} for the differential  rate of the process (\ref{eq:FeynDiagr-1}) in the form
%
%As a consequence of two separate volumes given by the source and detector, a product of two delta functions representing
%momentum conservation associated with both weak vertices appear in the last line of Eq. (\ref{ampl2}). 
%
%The differential decay-rate can be written as
\begin{eqnarray}
\label{eq:Master-1}
\nonumber
&&  d\Gamma^{\alpha\beta} (L) = 
%\delta(E_\beta + E'_D - E_D + E_\alpha + E'_S - E_S)\times\nonumber\\
 \sum_{k m} 
U_{\alpha k} U^*_{\beta k} U_{\alpha m} U^*_{\beta m}~\frac{e^{i (p_k-p_{m}) L}}{4\pi L^{2}}\times  \mathcal{F}^{\alpha\beta}_{km}\\
%
%\nonumber
%
%\nonumber
&&\delta(\mathbf{p}_k + \mathbf{p}_\alpha + \mathbf{p}'_S - \mathbf{p}_S)
\delta(\mathbf{p}_\beta + \mathbf{p}'_D - \mathbf{p}_D - \mathbf{p}_m)  \\
&&\frac{(2\pi)^7}{4 E_S E_D}~\delta(E_\beta + E'_D - E_D + E_\alpha + E'_S - E_S)\times \nonumber\\
%\nonumber
%&&  \delta(\mathbf{p}_k + \mathbf{p}_\alpha + \mathbf{p}'_S - \mathbf{p}_S)
%\delta(\mathbf{p}_\beta + \mathbf{p}'_D - \mathbf{p}_D - \mathbf{p}_m) 
%\times  
%\nonumber\\
%  && \sum_{\rm spin}
%%
%\frac{1}{2}\left(T_{k}^{\alpha\beta}\left(T_{m}^{\alpha\beta}\right)^{*}+ T_{m}^{\alpha\beta}\left(T_{k}^{\alpha\beta}\right)^{*}\right)
%%
%%  \frac{1}{2}\left(\langle f|T_k|i\rangle \langle f|T_m|i\rangle^* + \langle f|T_m|i\rangle \langle f|T_k|i\rangle^*\right)
  %\times
\nonumber
&&  \frac{1}{\hat{J}_S \hat{J}_D}\frac{d\mathbf{p}_\alpha}{2E_\alpha~(2\pi)^3}~\frac{d\mathbf{p}_\beta}{2E_\beta~(2\pi)^3}~
  \frac{d\mathbf{p'}_S}{2E'_S~(2\pi)^3} ~ \frac{d\mathbf{p}'_D}{2E'^{}_D~(2\pi)^3}.
\end{eqnarray}  
where
\begin{eqnarray}\label{eq:F-def}
\nonumber
\mathcal{F}^{\alpha\beta}_{km} &=&
%4\pi \delta(\mathbf{p}_k + \mathbf{p}_\alpha + \mathbf{p}'_S - \mathbf{p}_S)
%%\times \\
%%\nonumber
%%&&
%\delta(\mathbf{p}_\beta + \mathbf{p}'_D - \mathbf{p}_D - \mathbf{p}_m) \times \\
%\nonumber
 4\pi\sum_{\rm spin}
\frac{1}{2}\left(T_{k}^{\alpha\beta}\left(T_{m}^{\alpha\beta}\right)^{*}+ 
T_{m}^{\alpha\beta}\left(T_{k}^{\alpha\beta}\right)^{*}\right)
\end{eqnarray}
The factor $1/(\hat{J}_S\hat{J}_D)$ ($\hat{J}=2 J +1$) is due to averaging over spin projections of the initial hadrons
$N_S(P_S)$ and $N_D(P_D)$. 
%{\color{blue} 
Our master formula (\ref{eq:Master-1}) has been derived by applying the standard approach, which implies confining the system to the finite volume. In our case this is the volume 
$V_{S}+V_{D}$.  Therefore, we normalize the states of the particles  
$\Phi^{S}=S,S',\ell^{+}_{\alpha}$ and 
$\Phi^{D}=D,D',\ell^{-}_{\beta}$ as
%
%in the sub-volumes $V_{S}$ and $V_{D}$, respectively. 
\begin{eqnarray}\label{eq:normalization-1}
\langle \Phi^{S,D}(\mathbf{P}_{i})| \Phi^{S,D}(\mathbf{P}_{k})\rangle = (2\pi)^{3} 2 E_{k} \delta^{3}_{V_{S,D}} (\mathbf{P}_{i}-\mathbf{P}_{k})
\end{eqnarray}
and used the approximation 
\begin{eqnarray}\label{eq:Delta-V-app}
&&\delta^{3}_{V}(\mathbf{Q}_{n}-\mathbf{P})\delta^{3}_{V}(\mathbf{Q}_{m}-\mathbf{P}) \simeq \nonumber\\
&&~~~~~
  \frac{V}{(2\pi)^{3}} ~\frac{1}{2}\left(\delta^{3}_{V}(\mathbf{Q}_{n}-\mathbf{P}) + \delta^{3}_{V}(\mathbf{Q}_{m}-\mathbf{P})\right)\, .
\end{eqnarray}
%}
This is an exact relation for $n=m$.  For $n\neq m$ it is still a good approximation, since the neutrino square mass differences $m^{2}_{n}-m^{2}_{m}$, according to the neutrino oscillation data, are very small compared to the typical energies of the particles involved in the process (\ref{eq:FeynDiagr-1}).
%{\color{blue} 
These two aspects (\ref{eq:normalization-1}) and (\ref{eq:Delta-V-app}) of our approach  are crucial for the cancellation of all the volume factors in Eq.~(\ref{eq:Master-1}), making our master formula a well-defined physical quantity.
%This master formula for the differential is the central result of our approach. 

%{\color{blue}
One of the important messages we have from this formula is that in QFT there is no factorization of the rate (\ref{eq:Master-1}) of the underlying process  (\ref{eq:FeynDiagr-1}) allowing one to devide  it in a sequence of three independent processes (\ref{eq:Three-step-1}) as  assumed in the standard QM approach.
%}
%in Eq. (\ref{eq:def-T-1}) as QFT neutrino propagator is introduced. The factorization can not be achieved by
%a transformation
%\begin{eqnarray}
%&&(1-\gamma_5)~ \sla{Q}_k = \\
%&&\frac{1}{2}~ (1-\gamma_5)~ \sum_{\lambda_k} u(Q_k;\lambda_k) \overline{u}(Q_k;\lambda_k)~ 
%(1+\gamma_5)\nonumber
%\end{eqnarray}  
%as neutrinos are massive particles and there is a sum over both polarizations of neutrino.
%Consequently, the trace corresponding to the lepton current in Eq. (\ref{eq:def-T-1})
%has to be calculated.
%}

%%%%%%%---------------

%We note that the transition probability is final and does not depend on the volumes of
%source and detector ($V_S$ and $V_D$). 
%For a product of two delta functions related to the momentum conservation
%due to first or second weak interactions we used
%\begin{eqnarray}
%&&  \delta(\mathbf{p}_{k, m} - \mathbf{p})  \int_{V_{S,D}} e^{i(\mathbf{p}_{m,k} - \mathbf{p})\cdot\mathbf{x}_{S,D}}~ d \mathbf{x}_{S,D}\nonumber\\
%&&~~~~~~~~    = \lim_{V_{S,D}\rightarrow \infty} \delta(\mathbf{p}_{k, m} - \mathbf{p})~ V_{S,D}.
%\end{eqnarray}
%This relation is exact for $k=m$. For $k\ne m$ it works well as in the case of neutrino oscillations
%$|m^2_k - m^2_l|/E_\nu$ is a very small quantity. 

For illustrative purposes, we consider a concrete version of the generic process (\ref{eq:FeynDiagr-1}). One of the simplest cases is the following
\begin{eqnarray}\label{eq:FeynDiagr-2-1}     
\pi^{+} + n \to \mu^{+} + e^{-} + p
\end{eqnarray}
with $\mu^{+}$ and $e^{-}$ emitted in the source {\bf S} and detector {\bf D}, respectively.

{\it In the conventional approach}, based on the concept of the quantum-mechanical oscillation probability (\ref{eq:P-oscill-1}) the process (\ref{eq:FeynDiagr-1}) is factorized in the three independent processes 
%}

%eq:Detection-1 eq:Emission-1
%
%the situation where $\nu_\mu$ is produced from $\pi^+$ decay at a source,
%and at a later time, the neutrino will hit a neutron target to produce $e^-$.
\begin{eqnarray}
  \pi^+ \rightarrow \mu^+ + \nu_\mu, ~~~
  \nu_\mu \rightarrow \nu_e, ~~~
  \nu_e + n \rightarrow p + e^-.
  \label{pibeta}
\end{eqnarray}
%{\color{blue}
This set of the processes underlies the observed  atmospheric neutrino oscillations with the target neutron belonging to the detector target nucleus.
For simplicity, we do not take into account nuclear effects and assume that the kinetic energy, $E_{\pi}$, of the decaying pion, measured in the rest frame of the target neutron, is negligibly small compared to its mass $m_{\pi}$.
%
%For sake of simplicity, the kinetic energy of decaying pion is assumed to be negligible when
%compared to its mass in the frame associated with the decaying nucleon.
%} 
The production rate of this compound process,
according to 
%following the standard approach (see 
Eq.~(\ref{prodrnn}), is
\begin{eqnarray}
  \Gamma_{osc}^{\pi^+ n}  &=& \int \frac{d \Phi_\nu(E_\nu)}{d E_\nu}~
  \frac{P_{{\nu}_\mu {\nu}_e}(E_\nu)}{4\pi L^2}~ \sigma(E_\nu) ~dE_\nu\nonumber\\
&=& \frac{1}{2\pi^2}~G^2_\beta ~\left(\frac{f_\pi}{\sqrt{2}}\right)^2
  ~ \frac{m^2_\mu}{m_\pi} ~ E^2_\nu\times \nonumber\\
&&~~~\frac{P_{{\nu}_\mu {\nu}_e}(E_\nu)}{4\pi L^2}~\left(g^2_V + 3 g^2_A \right)~p_e E_e. 
\label{standardpr}
\end{eqnarray}
Here, $G_\beta = G_F \cos{\theta_C}$, where $\cos{\theta_C}$ is the Cabbibo angle. 
%{\color{orange} 
$E_\nu$ ($p_\nu$) and $E_e$ ($p_e$) are  the energies (momenta) of neutrino and electron,
respectively. In the above formula we used an approximation $E_\nu \simeq E_e$ in the rest frame of the decaying pion and the target neutron, where 
$E_\nu = m_\pi (1 - m^2_\mu/m^2_\pi)/2$ with  
$m_\pi$ ($m_\mu$) being the mass of pion (muon). Vector
and axial-vector coupling constants of the nucleon are denoted by  and $g_V$ and $g_A$ are, respectively. As usual $f_\pi$ is the pion decay constant.
%}

For the neutrino energy distribution and the cross-section  we used in Eq.~(\ref{prodrnn})  the well-known expressions
%\cite{Spectrum}
%In the production and absorption processes of neutrinos the neutrino mass is neglected. We have 
\begin{eqnarray}
  \frac{d \Phi_\nu(E_\nu)}{d E_\nu} &=& 
  \frac{1}{2\pi} G^2_\beta ~\left(\frac{f_\pi}{\sqrt{2}}\right)^2
 ~\frac{m^2_\mu}{m_\pi} ~E^2_\nu\, , \\
\nonumber  
  \sigma(E_\nu) &=& \frac{1}{\pi}~ G^2_\beta~\left( g^2_V + 3 g^2_A \right)~p_e E_e. 
\end{eqnarray}
%{\color{blue}

{\it In our QFT approach}, we apply Eq.~(\ref{eq:Master-1})  for  the process (\ref{eq:FeynDiagr-2-1}).
In this case we have
%
%the QFT approach for the differential decay-rate of the sequence of weak processes
%in (\ref{pibeta}) we find
%\begin{eqnarray}
%&&  d\Gamma^{\pi^+ n} =\nonumber\\
%&&\frac{1}{4 E_\pi E_N}~(2\pi)^7 \delta(E_e + E_p - E_n + E_\mu - E_\pi)\times\nonumber\\
%&& \sum_{k m} 
%  \delta(\mathbf{p}_k + \mathbf{p}_\mu - \mathbf{p}_\pi)
%~\delta(\mathbf{p}_e + \mathbf{p}_p - \mathbf{p}_n - \mathbf{p}_m) \times  \nonumber\\
%  && \frac{1}{2} \sum_{\rm spin}
%  \frac{1}{2}\left(\langle f|T_k|i\rangle \langle f|T_m|i\rangle^* + \langle f|T_m|i\rangle \langle f|T_k|i\rangle^*\right)
%  \times\nonumber\\
%&& \frac{d\mathbf{p}_e}{2E_e~(2\pi)^3}~\frac{d\mathbf{p}_\mu}{2E_\mu~(2\pi)^3}~
%  \frac{d\mathbf{p}_n}{2E^{}_n~(2\pi)^3}.\nonumber\\
%\end{eqnarray}  
%where
\begin{eqnarray}  
&& T^{\mu e}_{k} =
%\langle f|T_k|i\rangle = 
%e^{-i(\mathbf{p}_e + \mathbf{p}_p - \mathbf{p}^{}_n)\cdot \mathbf{L}}~
  J^\rho_{S}(P^{}_{\pi}) J^\sigma_{D}(P_{p},P^{}_{n})
  %~\frac{e^{i p_k L}}{4\pi L}
  \times\nonumber\\
&& U_{\mu k} U^*_{e k}~\overline{v}(P_\mu;\lambda_{\mu})\gamma_\rho(1-\gamma_5)  \sla{Q}_k
\gamma_\sigma u(P_e;\lambda_{e})
\end{eqnarray}
with 
\begin{eqnarray}  
  J^\rho_{S}(P^{}_{\pi}) &=& \frac{G_\beta}{\sqrt{2}} i f_\pi ({P}_\pi)^\mu~\nonumber\\
  J^\sigma_{D}(P_{p},P^{}_{n}) &=& \frac{G_\beta}{\sqrt{2}} \overline{u}(P_p) \gamma^\sigma (g_V - g_A \gamma_5) u({P}_n)
\end{eqnarray}
Here, $P_{\pi} \equiv (E_{\pi},\mathbf{p}_{\pi})$, $P_{n} \equiv (E_{n},\mathbf{p}_{n})$, 
$P_{p} \equiv (E_{p},\mathbf{p}_{p})$, $P_{\mu} \equiv (E_{\mu},\mathbf{p}_{\mu})$,
$P_{e} \equiv (E_{e},\mathbf{p}_{e})$, and $P_{k} \equiv (E_{\nu},\mathbf{p}_{k})$ with
$E_\nu = E_\pi - E_\mu = E_e + E_p - E_n$.
%
%The calculation of the decay-rate $d\Gamma^{\pi^+ n}$ involves 
%the evaluation of product of traces and integration over angles of emitted electron.
%We have
%\begin{eqnarray}
%  &&\int d\Omega_e ~{\rm Tr}\left(\sla{P}_e + m_e\right) \gamma^\sigma \sla{P}_k P_L
%\left(\sla{P}_\mu - m_\mu\right) \sla{P}_m \gamma^\delta P_R \times \nonumber\\  
%&&{\rm Tr } \left(\sla{P_p}+ m_p\right) \gamma_\sigma (g_V - g_A \gamma_5) \left(\sla{P_n}+ m_n\right)
%\gamma_\delta (g_V - g_A \gamma_5)\nonumber\\
%&\simeq& 4\pi~32~ m^2_N~\left( g^2_V + 3 g^2_A\right)~E_e E_\mu\left(E_\nu^2 + {p}_k {p}_m\right)
%\end{eqnarray}
%with $P_{L,R} = (1\mp\gamma_5)/2$. 
%
%The rest frame of pion at the rest $\mathbf{p}_\pi =0$
%is considered. Only the dominant terms proportional to $m_n m_p $ or $E_n E_p$ are kept
%($E_{n,p}\simeq m_{n,p} = m_N$). In addition, terms suppressed by the factor
%$E_e/E_\mu \simeq E_\nu/E_\mu \simeq (m_\pi^2 - m^2_\mu)/(m_\pi^2 + m^2_\mu)$ are neglected.
%We note that $\mathbf{p}_k\cdot\mathbf{p}_m = p_k p_m$. 
%
After integration over the phase space in (\ref{eq:Master-1})  and averaging the initial neutron polarization as well as summing over the polarizations of the final particles we find the total  rate 
of the process (\ref{eq:FeynDiagr-2-1}) in the form, which differs from that in the conventional approach (\ref{standardpr}). This is because in the QFT this process is not factorizable in the three independent ones shown in  (\ref{pibeta}). 
%{\color{blue}
Nevertheless, this happens under certain reasonable approximations, so  that in the rest frame of the target neutron
 we arrive at
 %}
%
\begin{eqnarray}\label{decratfin} 
\Gamma_{QFT}^{\pi^+ n} &=& \frac{1}{2\pi^2}~G^2_\beta ~\left(\frac{f_\pi}{\sqrt{2}}\right)^2
    ~ \frac{m^2_\mu}{m_\pi} ~[g^2_V + 3 g^2_A ]~~p_e E_e \\
    \nonumber
&&    \sum_{k m} U_{ek} U^*_{\mu k}~U^*_{ek} U_{\mu k}~
    \frac{e^{i(p_m-p_k) L}}{4\pi L^2}~\frac{\left(E^2_\nu + {p}_k {p}_m\right)}{2}. 
\end{eqnarray}
%{\color{blue} 
%Here, the rest frame of the target
%associated with the decaying 
%neutron is assumed. 
From Eq. (\ref{nuencon})
it follows that $E_e \simeq E_\nu $ as $E_{n,p}\simeq m_{n,p} = m_N$. Only dominant terms proportional
to $m_n m_p $ or $E_n E_p$ are retained. In addition, terms suppressed by the factor
$E_e/E_\mu \simeq E_\nu/E_\mu \simeq (m_\pi^2 - m^2_\mu)/(m_\pi^2 + m^2_\mu)$ are neglected.
%} 
%We note that $\mathbf{p}_k\cdot\mathbf{p}_m = p_k p_m$. 
%
%
%The rest frame of pion at the rest $\mathbf{p}_\pi =0$
%is considered. Only the dominant terms proportional to $m_n m_p $ or $E_n E_p$ are kept
%($E_{n,p}\simeq m_{n,p} = m_N$). In addition, terms suppressed by the factor
%$E_e/E_\mu \simeq E_\nu/E_\mu \simeq (m_\pi^2 - m^2_\mu)/(m_\pi^2 + m^2_\mu)$ are neglected.
%We note that $\mathbf{p}_k\cdot\mathbf{p}_m = p_k p_m$. 
%

Now we compare  $\Gamma_{QFT}^{\pi^+ n}$ in (\ref{decratfin}) with  
$\Gamma_{osc}^{\pi^+ n}$ in (\ref{standardpr}). They both give the rate of the same process (\ref{eq:FeynDiagr-2-1}), while the conventional approach (\ref{standardpr}) assumes factorization of the process (\ref{eq:FeynDiagr-2-1}) in the three separate steps (\ref{pibeta}).  
%Under the well justified approximations $p_{k,m} \simeq E_\nu$  and $m_{k,m}/E_\nu \ll 1$ in
%(\ref{decratfin}) 
We find that both results coincide in structure if we interpret  as the QFT oscillation probability the expression
\begin{eqnarray}
\label{eq:QFT-Probability-1}
\mathcal{P}^{QFT}_{\alpha\beta}&=&
\frac{1}{2}\sum_{k m} U_{ek} U^*_{\mu k}~U^*_{ek} U_{\mu k}~
e^{i(p_m-p_k) L}\times \\
\nonumber
&&~\times 
%{\color{blue}
\left(1 + \frac{{p}_k {p}_m}{E^{2}_{\nu}}\right)
%}
%\nonumber\\
%&&\simeq \frac{P_{{\nu}_\mu {\nu}_e}(E_\nu)}{4\pi L^2}
%
\end{eqnarray}
%
 %By comparing $\Gamma_{QFT}^{\pi^+ n}$ in (\ref{decratfin}) with  $\Gamma_{osc}^{\pi^+ n}$ in (\ref{standardpr})
%we conclude that the  results of the conventional and our QFT  approaches do agree with each other
%in the framework of 
%{\color{blue}
For $m_{k,m}/E_\nu \ll 1$ this expression coincides with the conventional {\color{blue} QM} formula for the neutrino oscillation probability (\ref{eq:P-oscill-1}) providing to it small corrections.  Note that our formula (\ref{eq:QFT-Probability-1})  is valid for  the case of heavy neutrinos, when $M_{k,m}/E_\nu \sim 1$, if  $M^{2}_{i}-M^{2}_{k}\ll E^{2}_{\nu}$.  The latter condition is required in our approach in order to guarantee reliability of 
(\ref{eq:Delta-V-app}).  Both of these conditions are met in neutrino mass models with an extended neutrino spectrum, including light active neutrinos and a few quasi-degenerate heavy neutral leptons. An example of this kind of model is given by the inverse seesaw mechanism.  
%
%{\color{blue}
Interesting consequences of quasi-degenerate heavy neutrino oscillations in particle decays, such as B-meson, tau, Higgs and W,   have been recently revealed in Refs.~\cite{Cvetic:2015ura,Zamora-Saa:2016ito,Tapia:2021gne,Cvetic:2021itw}, where the conventional approach to neutrino oscillations had been applied  and the plausibility of searching for their manifestations in various experiments was argued. Note, that our approach is not applicable to this type of oscillations searched for under the experimental conditions incompatible with  our assumption about long-baseline settings {\bf (a)-(d)}. We postpone the corresponding extensions of our analysis, including this and some other settings, for future publications.
%}
%Recall, that by performing the Taylor expansion over this parameter for the expression $i(p_m-p_k) L$
%in exponential function,  we get
%\begin{eqnarray}
%&&\sum_{k m} U_{ek} U^*_{\mu k}~U^*_{ek} U_{\mu k}~
%\frac{e^{i(p_m-p_k) L}}{4\pi L^2}~\frac{\left(E^2_\nu + {p}_k {p}_m\right)}{2}\nonumber\\
%&&\simeq \frac{P_{{\nu}_\mu {\nu}_e}(E_\nu)}{4\pi L^2}.\nonumber\\
%\end{eqnarray}

In summary, we proposed a quantum field theory approach to neutrino oscillations in vacuum.
We applied the usual S-matrix formalism for the process (\ref{eq:FeynDiagr-1}) described by the diagram in Fig.~\ref{fig:feynman}. 
%We assumed that 
This process is mediated by
%, as a particular case, 
intermediate neutrinos, described by the usual propagators for the neutrino mass eigenstates.  The effect of neutrino oscillations on the way from one vertex to the other
is automatically incorporated in the propagator due to the PMNS mixing present in the vertices.
One of the key points of our approach is the election of the space-time setup {\bf (a)}-{\bf (d)} typical for the neutrino oscillation experiments.  Then we analyzed the space-time and analytic properties of the general S-matrix element (\ref{ampl}) for this process and obtained its form 
(\ref{ampl2}) adopted to the setup {\bf (a)}-{\bf (d)}.  With this at hand and properly normalizing the initial and final states (\ref{eq:normalization-1}) we derived our flavor conversion Master Formula for differential rate of the process (\ref{eq:FeynDiagr-1}) valid under the approximation $m^{2}_{i}-m^{2}_{k}\ll E_{\nu}$.  
For the particular process (\ref{eq:FeynDiagr-2-1}) we demonstrated that our approach leads to a probability formula for $\nu_{\mu}\to \nu_{e}$ neutrino oscillation corrected with respect to the conventional one by  terms $\sim m_{k}/E_{\nu}$, which can be large for heavy neutrino quasi degenerate states.

The final point to be emphasized is that the concept of  the neutrino oscillation probability is not congruent with QFT and can only be introduced under certain assumptions, which, in principle, depend on the choice of particular process (\ref{eq:FeynDiagr-1}).
Therefore, in our opinion, the most safe approach to the analysis of neutrino oscillation data would be the direct use of the Master Formula (\ref{eq:Master-1}) for  extraction the oscillation parameters. 

\section*{Acknowledgments}

F.\v{S}. acknowledges support by the VEGA Grant Agency of the Slovak Republic under Contract
No. 1/0607/20 and by the Ministry of Education, Youth and Sports of the Czech
Republic under the INAFYM Grant No. CZ.02.1.01/0.0/0.0/16\_019/0000766. 
S.K. has been supported by ANID PIA/APOYO AFB220004 (Chile), ANID FONDECYT (Chile) No. 1190845 and ANID Programa Milenio code ICN2019\_044.

\end{document}